# Directed factor graph based fault diagnosis model construction for mode switching satellite power system

**Xiaolei Zhang. \* Yi Shen. \* Zhenhua Wang. \***

*\*School of Astronautics, Harbin Institute of Technology,*
*Harbin Institute of Technology,*
*Harbin, China, 15001, (e-mail: xiaoleizhang@hit.edu.cn)*
*\*(e-mail: shen@hit.edu.cn)*
*\*(e-mail: zhwang1987@gmail.com)*

**Abstract:** Satellite power system is a complex, highly interconnected hybrid system that exhibit nonlinear and mode switching behaviors. Directed factor graph is an inference model for fault diagnosis using probabilistic reasoning techniques. A novel approach for constructing the directed factor graph structure based on hybrid bond graph model is proposed. The system components status and their fault symptoms are treated as hypothesis and evidences respectively. The cause-effect relations between hypothesis and evidences are identified and concluded though qualitative equations and causal path analysis on hybrid bond graph model. A power supply module of a satellite power system is provided as case study to show the feasibility and validity of the proposed method.

## 1. INTRODUCTION

Because of the increasing complexity and unacceptable loss cost of the spacecraft, there is a need for the spacecraft to have ability to determine its subsystem and component health status. This required such complex engineering system possess the functions of fault detection and isolation (FDI). Many existing FDI methods based on mathematical model, which utilizing analytical redundancy relations to accomplish the fault diagnosis task, can generally be classified into model-based approach (see e.g. Isermann, 2005).

To detect and localize the faulty components in a system is the goal of the fault diagnosis. The model-based fault diagnosis incorporated structure information and cause-effect relationship about the system components. For the complex nonlinear spacecraft system with operational mode switching, the main problem of this methodology is that it is incapable of coping with the uncertainty of the system brought by the space severe environments and mode switching; also it can not utilize the experience knowledge about the system which acquired from experts and skilled operators.

In recent years, the probabilistic graph model (PGM) such as Bayesian Networks (BNs) based fault diagnosis techniques have been adopted to deal with uncertain information for system health monitoring (see e.g. Lerner et al., 2000; Verron et al., 2007; Yongli et al., 2006). The fundamental principle of PGM based fault diagnosis method is to calculate the $P(X|E=e)$, i.e. the probability distribution over some random variables $X$ given some evidence $E=e$, which means to query the probability of a certain component's fault given some detected fault symptoms.

The aim of the probabilistic graph model based method for system fault diagnosis is to express the cause-effect relations and the structure information about the system's components, in the mean time, to eliminate the model imprecision caused by system uncertainty existing in the traditional model-based method. But the BNs has its limitation in modelling ability for expressing cause-effect relations (see e.g. Frey, 2003). To overcome this drawback, a more powerful probabilistic graph model, the factor graph with directed edges – directed factor graph (DFG) (see e.g. Loeliger, 2004 and Frey, 2003) can be adopted to model the system cause-effect relations.

Directly constructing the probabilistic graph model from physical system's components for diagnosis purpose often need complicated computation (see e.g. Sahin et al., 2007) or based on existing diagnosis method (see e.g. Przytula & Milford, 2006). In this paper, hybrid bond graph (HBG) (see e.g. Mosterman, 1998) is used as the skeleton for constructing the DFG. The HBG has basic junctions for modelling the operational mode changing; the cause-effect relations can be set up through the causality of the HBG. The goal of constructing DFG for fault diagnosis could be accomplished through identifying random variables which represent the fault hypothesis and symptom information, determining directed edges linking nodes through the causality assignments of the HBG elements.

This paper is organized as follows. The knowledge of DFG is provided in Section 2. In Section 3, the construction of directed factor graph is formulated on the basis of hybrid bond graph model approach. Fault diagnosis mode construction procedure based on based on directed factor graph and its application to satellite power system (SPS) is presented in Section 4. Finally, the paper is concluded in Section 5.

## 2. DIRECTED FACTOR GRAPH

The directed factor graph is a kind of probabilistic graph model which can describe the problem environment, and then the diagnosis task is accomplished through reasoning which corresponds to probabilistic inference. After identifying all potentially relevant variables of a concerned system, the directed factor graph model describes how these variables can interact. This is achieved using the joint probability distribution of all the variables, typically corresponding to assumptions of independence of variables.

The basic elements of DFG are nodes and edges. Unlike Bayesian Network, the nodes in directed factor graph can represent random variables and functions. Each function depends on a subset of variables, and the function node is connected to the node corresponding to the subset of variables. The directed edges of DFG model express conditional distributions which represent cause-effect relation between variables and functions. In general, the directed factor graph $G=(X, F, E)$ consists of variable nodes $X=\{x_0, x_1, ..., x_n\}$, factor nodes $F=\{g_1, g_2, ..., g_m\}$ and directed edges $E$. The joint probability distribution is given by

$$f(x_0, x_1, ..., x_n) = \prod_{j=1}^{m} g_j(x_i, ..., x_k) \quad (1)$$

where $(x_0, x_1, ..., x_n) \subset (x_0, x_1, ..., x_n)$ denote the subset of variables that are connected to the function node. The edges $E$ with direction indicate the cause-effect relations which existed in the function and variables.

The representation framework of directed factor graph (DFG) allows reasoning under uncertainty. Component failure probability can be computed through probability inference mechanism. The edges of the DFG represent the cause-effect relation between the function node, and the function nodes of the directed factor graph are hierarchical representation of the structure information about the system which is comprised of different type of components.

Fig.1 (a) is a directed factor graph that represents a direct energy transfer (DET) power supply module of satellite power system which is shown in Fig.1 (b).

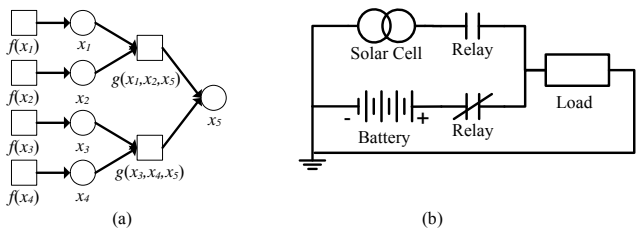

Fig.1. (a) directed factor graph for (b) corresponding DET power supply module of satellite power system

In Fig.1, the variable $x_1$ and $x_4$ denote the output current of solar cell and battery, $x_2$ and $x_3$ corresponding to the on/off status of relays connected to the solar cell and battery, the actual load input current is represented by variable $x_5$.

The joint distribution function represented by the DFG in Fig.1 (a) is formulated by

$$\begin{aligned} f(x_5, x_1, x_2, x_3, x_4) &= f(x_5 | x_1, x_2, x_3, x_4) f(x_1) f(x_2) f(x_3) f(x_4) \\ &= f(x_5 | x_1)^{x_2} f(x_5 | x_3)^{x_4} f(x_1) f(x_2) f(x_3) f(x_4) \\ &= g_1(x_5, x_1, x_2) g_2(x_5, x_3, x_4) g_3(x_1) g_4(x_2) g_5(x_3) g_6(x_4) \end{aligned} \quad (2)$$

where $g_i(x)$, $i=1,2,...,6$ denote the function nodes, $x$ is variables subset vector and $g_i(x)$ is listed in the following.

$$\begin{cases} g_1(x_5, x_1, x_2) = f(x_5 | x_1)^{x_2} \\ g_2(x_5, x_3, x_4) = f(x_5 | x_3)^{x_4} \\ g_3(x_1) = f(x_1) \quad g_4(x_2) = f(x_2) \\ g_5(x_3) = f(x_3) \quad g_6(x_4) = f(x_4) \end{cases} \quad (3)$$

Using probabilistic inference over the DFG for fault diagnosis purpose is typically a task of computing the probability of each node when other nodes' values are known. That means once some evidence about variables' states are determined, the effect of evidences could be propagated through the DFG. Specific to the DFG shown in Fig.1 (a), suppose that the load input current $x_5$ is noticed to exceed the threshold, we could use above principle to find out which component cause this fault. This means we need to calculate $f(x_1|x_5=\text{error})$, $f(x_2|x_5=\text{error})$, $f(x_3|x_5=\text{error})$ and $f(x_4|x_5=\text{error})$. Generally, this computing problem can be resolved through inference algorithm such as sum-product algorithm, EM algorithm, etc and each has its advantage over other in different situation (see e.g. Loeliger, 2004 and Frey, 2005).

## 3. BUILDING DIRECTED FACTOR GRAPH FOR FAULT DIAGNOSIS

The DFG based fault diagnosis method can overcome the uncertainty accompanied in the subsequent with nonlinear and operational mode switching existing in the satellite power system (SPS). The performance of fault diagnosis method will be greatly affected by the poor diagnosis model. To overcome this modelling problem, hybrid bond graph (HBG) (see e.g. Mosterman, 1998) is adopted to acting as the blueprint for generating the required DFG.

### 3.1 Hybrid bond graph

Bond graph (BG) (see e.g. Borutzky, 2010) is a modelling skeleton that makes use of basic elements to model the power exchange between the system components. In BG, the physical systems are classified into different basic elements associated with bonds, including two variables: effort, $e$, and flow, $f$; two power conservation junctions: serial 1-junction, and parallel, 0-junction; five physical primitives: Transformer, *TF* and Gyrator, *GY*, Inductance, *I*, Capacitance, *C* and Resistance, *R*; two ideal power source of flow, *Sf*, and effort *Se*.

The product of effort and flow defines the energy transfer from one element to another through bonds and junctions. In the electrical domain, effort and flow map to voltage and current respectively. On serial 1-junction, $f$ values are equal and $\Sigma e=0$, and correspondingly $e$ values are equal and $\Sigma f=0$ on parallel 0-junction. The behaviours of component are modelled by $R$, $C$ and $I$ capture energy dissipation in the system. The power source $Sf$ and $Se$ model the energy flow into and out of the system. Also the flow and effort sensor elements, $Df$ and $De$ are introduced to characteristic feature of sensor without affecting the system. Nonlinearities of physical system are modelled by expressing system parameters as functions of system variables using those basic elements.

For the sake of modelling on or off mode of operation, the controlled junction is introduced in BG. This basic element extends the bond graph to hybrid bond graph (HBG) (see e.g. Mosterman, 1998), and act as ideal switch in the model. Consider example electrical circuit shown in Fig.2 (a). The circuit consists of dc source $Se$ with voltage $V$, resistors $R1$ and $R2$, capacitor $C1$ and $C2$, sensor $D_{e1}$ and $D_{e2}$ measure the voltages over capacitor $C1$ and $C2$. The switch, $Sw$, is modelled by an ideal switching 1-junction, representing a series connection that can be on or off.

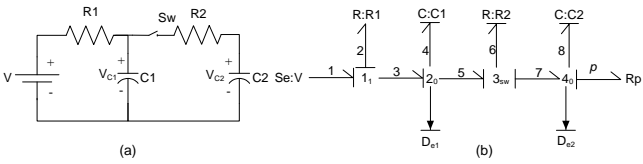

Fig.2. (a) circuit example with switching mode with (b) corresponding hybrid bond graph

In above hybrid bond graph shown in Fig.2 (b), an additional virtual resistor $Rp \approx \infty$ is used to obtain desirable unified computational causality (see e.g. Low, 2010) without losing the essence of the system's behaviour under the influence of the on/off system operation toggling.

### 3.2 Constructing directed factor graph

The structure of directed factor graph has some similarities to the hybrid bond graph model. In this subsection, the circuit example shown in Fig. 2 (a) is used as an illustration of construction procedure.

The goal of fault is achieved through monitoring the operational status of the system. The hypothesis variables which represent the physical component status should denote power source and physical primitives, that is, the basic elements $R1$, $R2$, $C1$ and $C2$, the power source $V$, since in a diagnostic application, the status of a system component is of concern. The basic sensor elements, $D_{e1}$ and $D_{e2}$ can act as evidence variables since they represent the voltage of $C1$ and $C2$. As mentioned before, there is a controlled 1-junction $3_{sw}$ exists in demonstration circuit, so it should be represented by a random variable $c$.

Performing inference algorithm over directed factor graph must rely on its directed links. These directed edges denote the causal relationship between different nodes. Based on unified computational causality, the causal path can be determined as illustrating in Fig.2 (b). Causal path explicitly denotes the relation between system variables based on causality of hybrid bond graph. With the aid of causal path, the random variables acquired in above step can be divided into two classes, one represents causes, the other represent their effects, and the directed edge should link causes to their effect. For example, in Fig.2 (b), the flow $f_4$ on bond 4 should be the effect and the cause is physical primitive $C1$. As shown in Fig.2 (a), when the capacitor $C1$ is breakdown, the current passing through the capacitor may not within the normal range.

The causal relations can be summarized in (4) by behaviours equations, it can be used to generate causal paths of the Bond graph.

$$\begin{cases} Se: e_1 = V \\ R: e_2 = R1 \times f_2 \\ C: f_4 = C1 \times (de_4/dt) \\ R: f_6 = e_6/R2 \\ C: f_8 = C2 \times (de_8/dt) \end{cases} \quad (4)$$

$$\begin{cases} 1_1: e_3 = e_1 - e_2 \quad f_2 = f_3 \\ 2_0: f_3 = f_4 + c \times f_5 \quad e_4 = e_3 \quad c \times e_5 = e_3 \quad e_3 = D_{e1} \\ 3_{sw}: c \times e_7 = c \times e_5 - c \times e_6 \quad cf_5 = c \times f_6 \quad c \times f_7 = c \times f_6 \\ 4_0: c \times f_7 = f_8 \quad e_7 = D_{e2} \quad e_8 = c \times e_7 \end{cases} \quad (5)$$

Constitutive equations of serial, parallel and controlled junctions are listed in (5). The controlled junction's status in second and third one of (5) is represented by a variable $c$.

The causal paths (see e.g. Borutzky, 2010) which deduced from hybrid graph are listed in (6). In causal paths, the components related with hypothesis variables are regarded as begin, and the evidence variables which represented by sensor elements are treated as end. The path direction represents the fault propagation in the physical system. It should be point out the controlled junction $c$ is treated as the begin.

$$\begin{aligned} &V \to e_1 \to e_3 \to D_{e1} \\ &R_1 \to e_2 \to e_3 \to D_{e1} \\ &C_1 \to f_4 \to f_3 \to f_2 \to e_2 \to e_3 \to D_{e1} \quad (6) \\ &c \to e_6 \to R_2 \to f_6 \to f_7 \to f_p \to e_p \to D_{e2} \\ &c \to f_7 \to f_8 \to C_2 \to e_8 \to D_{e2} \end{aligned}$$

In order to reducing the inference algorithm complexity, the superfluous variables in (4) and (5) which not belong to the set of hypothesis and evidence variables should be eliminated as many as possible with help of.

After above elimination procedure, the extra variables such as $f_3$ and $f_7$ are introduced as mediate variables to express the causal relations that are shown in Fig.2. These mediate variables can help to construct proper directed factor functions with diagnostic properties.

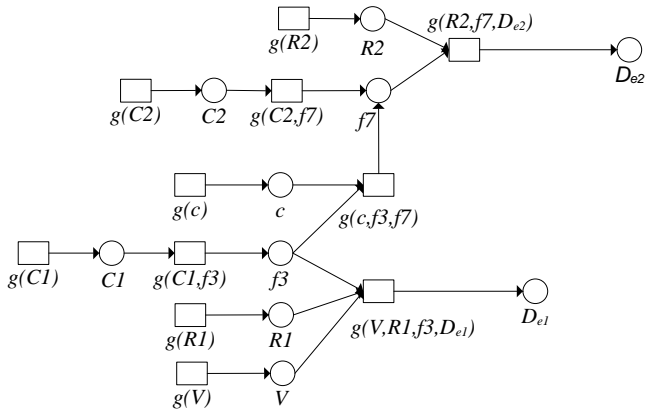

Fig.3. Directed factor graph for fault diagnosis

The above construction procedure of can be summarized as follows:

1) Build acausal hybrid bond graph for operational mode switching physical system, assign causality to acausal hybrid bond to get causal hybrid bond graph with the help of SCAPH algorithm and model approximation method (see e.g. Low, 2010).

2) Establish behaviour and constitutive equations of physical system components based on the unified causality assignment accomplished in step 1).

3) Identify the hypothesis events about physical systems components' status and evidences about fault symptoms; represent them into a set of random variables. The controlled junctions of hybrid bond graph should also be represented by random variables. Construct causal path between the hypothesis and symptom variables.

4) Eliminate superfluous variables in behaviour and constitutive equations which not belong to the set of hypothesis and evidence variables with the help of the causal paths analysis.

5) Preserve relevant mediate variables in causal paths which help to create causal links between variables and functions of the directed factor graph.

The fault diagnosis method based on directed factor graph combines the advantages of quality and quantity diagnosis methods. The directed links between variable and function nodes indicate that the function nodes are the consequence of the variable nodes through quality causal information. As describe above, the inference algorithms can use available data to figure out the fault probabilities of system components.

## 4. CASE STUDY

The satellite power system (SPS) provides the primary power source for all on-board systems. A typical satellite power system comprises a primary power source (solar array), an energy storage system (rechargeable batteries), shunt regulators, and so on. It contains components from multiple disciplines such as photovoltaic, electrochemistry, electrical system and power electronics.

Typically, the components in SPS could switch their operational mode based on the system requirements. As shown in Fig.4 (a), the components in dashed frame are two solar panels with shunt regulators and a rechargeable battery. These three main components and necessary auxiliary components make up a sequential shunt regulator power supply module. When the bus power is supplied by the solar panel under sun illumination, the bus voltage can be adjusted through the shunt regulator switching manipulation. The rechargeable battery begin to function when the sunlight are blocked from reaching the solar panels.

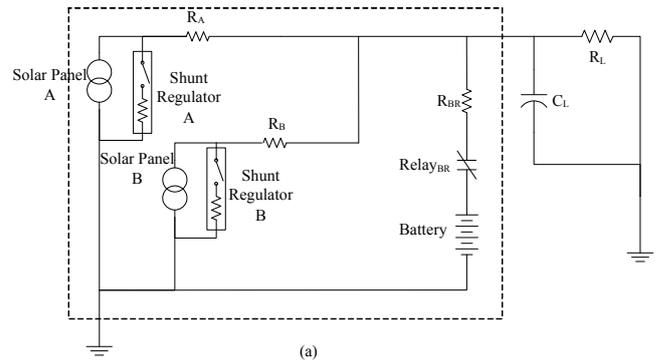

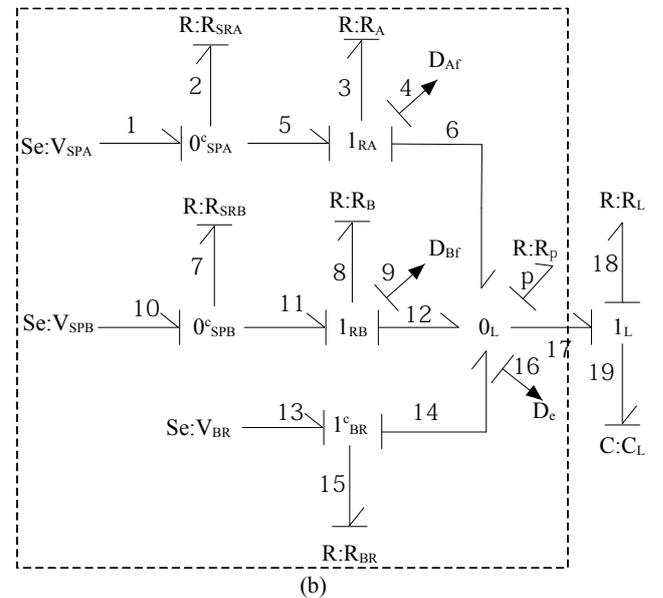

Fig.4 (a) Shunt regulator power supply module of satellite power system and (b) corresponding hybrid bond graph

As the first step to establish directed factor graph, in Fig.4 (b), the hybrid bond graph is constructed. The basic bond graph elements in dotted-line frame represent corresponding physical system components under different operation mode through ideal controlled switches. The ideal controlled 0-junctions $0^c_{SPA}$ and $0^c_{SPB}$ in Fig.4 (b) represent the switches in shunt regulators. The relay serial connected with the rechargeable battery is denoted by controlled 1-junction $1^c_{BR}$.

The effect source $V_{SPA}$, $V_{SPB}$ and $V_{BR}$ the solar panel A, B and the battery respectively. Two flow sensors, $D_{Af}$ and $D_{Bf}$, one effect sensor $D_e$, are selected from hybrid bond graph, denote the current of the solar panels and the voltage of the shunt regulator power supply module. The causal strokes assignment also has been accomplished through SCAPH algorithm and model approximation method (see e.g. Low, 2010).

For simplicity, the elements behaviour equations and junction constitutive equations are omitted. Only the hypothesis events variables and fault symptoms acting as begin and end nodes are listed in causal path in (6). It is necessary to point out that the controlled junction variables in (6), $c_{SPA}$, $c_{SPB}$ and $c_{BR}$, are not at begin or end of the causal path, but they should also be treated as hypothesis variables.

$$\begin{aligned}
&V_{SPA} \to e_1 \to c_{SPA} \to e_5 \to e_3 \to R_A \to f_3 \to D_{Af} \\
&R_{SRA} \to f_2 \nearrow \qquad\qquad\qquad \searrow f_6 \to f_p \to R_p \to e_p \to D_e \\
&V_{SPB} \to e_{10} \to c_{SPB} \to e_{11} \to e_8 \to R_B \to f_8 \to D_{Bf} \\
&R_{SRB} \to f_7 \nearrow \qquad\qquad\qquad \searrow f_{12} \to f_p \to R_p \to e_p \to D_e \\
&V_{BR} \to e_{13} \to c_{BR} \to e_{15} \to R_{BR} \to f_{15} \to f_{14} \to f_p \to R_p \to e_p \to D_e
\end{aligned} \quad (6)$$

After elimination superfluous variables with help of causal path analysis, the mediate variables $f_3$ and $f_8$ which related with hypothesis variables are preserved. From the causal paths listed in (6), the directed links also can be established. In Fig.5, the directed factor graph for fault diagnosis is constructed based on above steps.

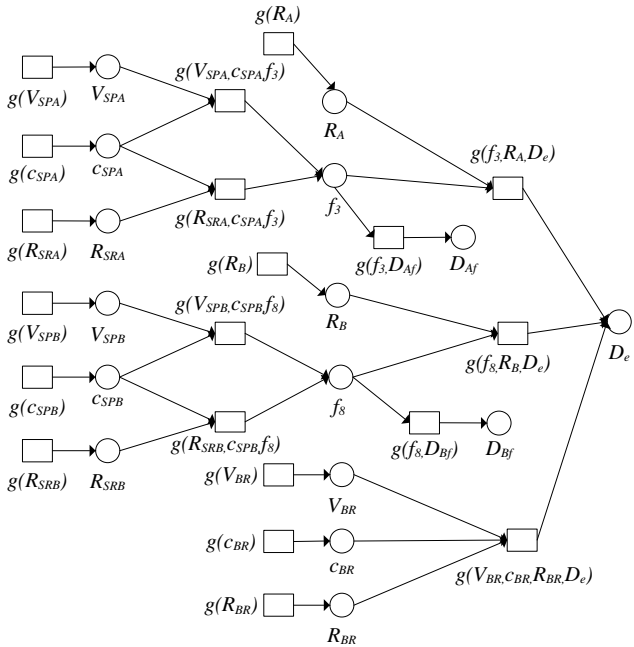

Fig.5 Diagnostic directed factor graph of satellite power system

Directed factor graph shown in Fig.5 represents the probabilistic relations between hypothesis events and fault symptoms, and supply the precondition for fault diagnosis under uncertain environments. The function nodes in directed factor graph can represent condition probabilistic functions of random variables. For example, the joint probabilistic function can be represented using function nodes of directed factor graph in (7).

$$\begin{aligned}
g(V_{SPA}, R_{SRA}, c_{SPA}, f_3) = &\, g(V_{SPA}) g(R_{SRA}) \cdot \\
&g(c_{SPA}) g(f_3) g(V_{SPA}, c_{SPA}, f_3) g(R_{SRA}, c_{SPA}, f_3)
\end{aligned} \quad (7)$$

More detail introduction about fault diagnosis using probabilistic inference techniques can be found in the references (see e.g. Lerner et al., 2000; Verron et al., 2007; Yongli et al., 2006).

## 5. CONCLUSIONS

The probabilistic inference techniques for fault diagnosis based on probabilistic graphical model such as directed factor graph have been utilized to deal with uncertain information in system health monitoring. But the diagnostic model structure of these techniques is often obtained by method of trial and error, through expert knowledge or time-consuming computation method. This contribution presents a novel approach on constructing a directed factor graph from hybrid bond graph model for mode switching hybrid system fault diagnosis. Fault symptoms and hypothesis variables are first identified, acting as the basis of diagnosis model. The causal links between variables and function nodes are generated from the set of qualitative behaviour and constitutive equations and causal path analysis on hybrid bond graph. The construction steps are explained in detail through a circuit example. Also this construction procedure is demonstrated by a case study using a mode switch satellite power system. This method pave the way for efficient and effective fault diagnosis under uncertain knowledge or incomplete information, and useful probabilistic inference algorithms can be developed for large complex hybrid systems..

## ACKNOLEDGEMENT

This work was supported by National Natural Science Foundation of China under the Project No. 60874054.